\documentclass{PoS}

\renewcommand{\r}{\mathbf{r}}
\renewcommand{\k}{\mathbf{k}}
\renewcommand{\d}{\mathrm{d}}
\newcommand{\Tr}{\widetilde{\mathrm{Tr}}}
\newcommand{\C}{\widetilde{C}}

\usepackage{amsmath}
\usepackage{amssymb}

\title{Ultraviolet divergences in the cyclic Wilson loop and their renormalization}

\ShortTitle{Renormalization of the cyclic Wilson loop}

\author{\speaker{Matthias  Berwein}
        \\
        Technische Universit\"at M\"unchen (Germany)\\
        E-mail: \email{matthias.berwein@mytum.de}}

\abstract{We discuss the cyclic Wilson loop, i.e. a rectangular Wilson loop that spans the entire compactified time axis in the imaginary time formalism. The result of a perturbative calculation at ${\cal O}(\alpha_s^2)$ is given, with the main focus on the ultraviolet divergences of this operator. Based on a general analysis of divergences in loop diagrams, we find that, unlike usual Wilson loops, the cyclic loop does not have cusp divergences but intersection divergences, where the intersections arise due to the periodic boundary conditions. As a result, the cyclic Wilson loop itself is not multiplicatively renormalizable, but the difference between it and the correlator of two Polyakov loops is.}

\FullConference{Xth Quark Confinement and the Hadron Spectrum,\\
		October 8-12, 2012\\
		TUM Campus Garching, Munich, Germany}

\begin{document}

\section{Introduction}
\label{1}

Wilson loops are very useful operators to study heavy quarkonium systems~\cite{Brambilla:2004wf,Brambilla:2004jw, Brambilla:2010cs}. In the infinite mass limit the propagator of a heavy quark can be expressed through a straight Wilson line in the time direction. Wilson lines also allow for the definition of gauge invariant states of two spatially separated quarks. Therefore the time evolution of a gauge invariant quark-antiquark state can be described by the expectation value of the trace of a rectangular Wilson loop. In the infinite time limit one can project out the lowest energy eigenvalue~\cite{Susskind:1976pi,Brown:1979ya}, and it's possible to account for finite mass corrections through insertions of gluonic operators into the Wilson loop~\cite{Brambilla:2004jw}. At finite temperature there is a similar operator called the \textit{Polyakov loop}, which consists of a straight Wilson line along the whole of the compactified time axis in the imaginary time formalism. The expectation value of the correlator of two traced Polyakov loops is related to the free energy of a system of an infinitely heavy quark and antiquark, and has a known spectral decomposition in terms of the energy eigenvalues of the static QCD Hamiltonian~\cite{Luscher:2002qv,Jahn:2004qr}. A Wilson loop, called the \textit{cyclic Wilson loop}, with two Polyakov loops as its time-like sides, is expected to have similar interpretations.

This paper is concerned with the properties of the cyclic Wilson loop, in particular its divergence structure. Typically, rectangular Wilson loops have ultraviolet divergences in addition to the usual ones related to charge renormalization. These additional divergences are called \textit{cusp divergences}, because they are related to the four right-angled corners of the loop's contour. They are known to factorize and can be removed by a multiplicative constant~\cite{Dotsenko:1979wb,Brandt:1981kf}. However, it was found that the additional divergences of the cyclic Wilson loop are different and cannot be removed by a multiplicative constant~\cite{Burnier:2009bk}. We present a solution to this problem by investigating the origins of UV divergences in terms of loop diagrams, and demonstrate that the difference between the cyclic Wilson loop and the correlator of two Polyakov loops is indeed multiplicatively renormalizable.

The paper is organized as follows. Section~\ref{2} will detail some properties of the cyclic Wilson loop, which will be instrumental in the following discussion, and give a perturbative calculation at~${\cal O}(\alpha_s^2)$. Section~\ref{3} will then discuss origins and different types of UV divergences in loop functions and apply this analysis to the case of the cyclic Wilson loop. Section~\ref{4} will then test the general statement of section~\ref{3} at ${\cal O}(\alpha_s^3)$, while section~\ref{5} contains our conclusions. All these results are explained in more detail in our paper~\cite{Berwein:2012mw}.

\section{Properties of the cyclic Wilson loop}
\label{2}

The Wilson line operator, for a line going from $y$ to $x$ along the contour $C$, is given by
\begin{equation}
 W(x,y;C)={\cal P}\exp\left[ig\int_{C(x,y)}\d z_\mu A_\mu(z)\right]\,,
\label{WLine}
\end{equation}
where ${\cal P}$ stands for path ordering of the colour matrices in $A_\mu(z)$ along the contour $C$. Since we will use the imaginary time formalism of finite temperature field theory, we work in Euclidean space with periodic boundary conditions in the time direction. Under a gauge transformation $V\in SU(N)$ Wilson lines transform as $V(x)W(x,y;C)V^\dagger(y)$, so that the trace of a Wilson line with a closed contour, i.e. a Wilson loop, is a gauge invariant operator.

A Polyakov loop operator is given by a straight Wilson line, which runs once around the compactified time axis:
\begin{equation}
 P(\r,T)={\cal P}\exp\left[ig\int_0^{1/T}\d\tau A_0(\tau,\r)\right]\,.
\end{equation}
Because of the periodic boundary conditions, the contour of a Polyakov loop is closed and the trace over a Polyakov loop is gauge invariant. The correlator of two Polyakov loops, hereafter simply called the \textit{Polyakov loop correlator}, is then given by
\begin{equation}
 P_c(r,T)=\left\langle\Tr[P^\dagger(0,T)]\Tr[P(\r,T)]\right\rangle\,,
\end{equation}
where $\Tr$ denotes the trace over colour matrices divided by the number of colours $N$. A single Polyakov loop operator does not depend on the position vector $\r$ because of translation invariance, but the correlator of two Polyakov loops does depend on their distance $r=|\r|$. $P_c$ is related to the free energy of an infinitely heavy quark and antiquark in a thermal medium.

The \textit{cyclic Wilson loop} is defined similarly to the Polyakov loop correlator, but instead of having two separate traces over the Polyakov loop operators, their endpoints are connected by two spatial Wilson lines to obtain a gauge invariant object. The cyclic Wilson loop therefore consists of four straight Wilson lines, the two corresponding to Polyakov loops will be called \textit{quark lines}, the two spatial lines will be called \textit{strings}.
\begin{equation}
 W_c(r,T)=\left\langle\Tr\left[W(\r,0)P^\dagger(0,T)W(0,\r)P(\r,T)\right]\right\rangle\,.
\label{CWL}
\end{equation}
In the arguments of the string operators only the spatial components are displayed, the temporal components are always $0$ or $1/T$, which is equivalent. Because Wilson lines are unitary operators, and because exchanging start and end point of a Wilson line corresponds to taking the hermitian conjugate, the two string operators are inverse to each other. This leads to many cancellations of Feynman diagrams in a perturbative expansion of the cyclic Wilson loop. Since this is a direct consequence of the periodic boundary conditions - in a usual rectangular Wilson loop the strings are separated in the time component and therefore not inverse to each other - we will call this effect \textit{cyclicity cancellation}.

A perturbative expansion of the cyclic Wilson loop gives Feynman diagrams, which consist of the rectangular contour specified in \eqref{CWL} and gluons attached to points on this contour. These endpoints of gluon propagators each give a factor of $ig$ and are integrated along the contour according to the path ordering prescription (cf.~eq.~\eqref{WLine}). We will refer to them as \textit{line vertices}, as opposed to \textit{internal vertices} connecting gluons to other gluon, ghost or quark propagators (cf.~ref.~\cite{Brandt:1982gz}). Now, cyclicity cancellation means that all diagrams cancel which have at least one line vertex on a string and no line vertex on at least one quark line.

Another useful feature of the cyclic Wilson loop (or in general any closed Wilson loop) is expressed in the exponentiation theorem \cite{Gatheral:1983cz,Frenkel:1984pz}, which says that the perturbative expansion in Feynman diagrams can be resummed and exponentiated:
\begin{equation}
 W_c=\sum_\gamma C(\gamma)W(\gamma)=\exp\left[\sum_{\gamma\in\mathrm{2PI}}\C(\gamma)W(\gamma)\right]\,,
\label{exp}
\end{equation}
where $\gamma$ denotes the Feynman diagrams, $C(\gamma)$ represents the colour factor of a diagram, which is given by the normalized trace over the colour matrices and contractions with structure constants from this diagram, and $W(\gamma)$ the value of a diagram without its colour factor, i.e.~the result of applying the Feynman rules to a diagram and performing the corresponding integrations. Eq.~\eqref{exp} states that the logarithm of the cyclic Wilson loop can be expressed as a sum of the same Feynman diagrams, but only the subset of \textit{two-particle irreducible} (2PI) diagrams contributes. A diagram is reducible, if the contour can be cut in two places such that the two pieces of the contour are not connected to each other by gluon, ghost or light quark propagators (excluding the case where there is no gluon attached to one of the pieces at all). If this is not possible, then it is a 2PI diagram. So the logarithm of $W_c$ is given by a series over only 2PI diagrams, however their colour factors have to be replaced by the \textit{colour-connected coefficients} of each diagram. These are denoted by $\C$ in eq.~\eqref{exp}. We refer to refs. \cite{Gatheral:1983cz,Frenkel:1984pz,Berwein:2012mw} for an explicit definition of these coefficients.

\begin{figure}
 \includegraphics[width=\linewidth]{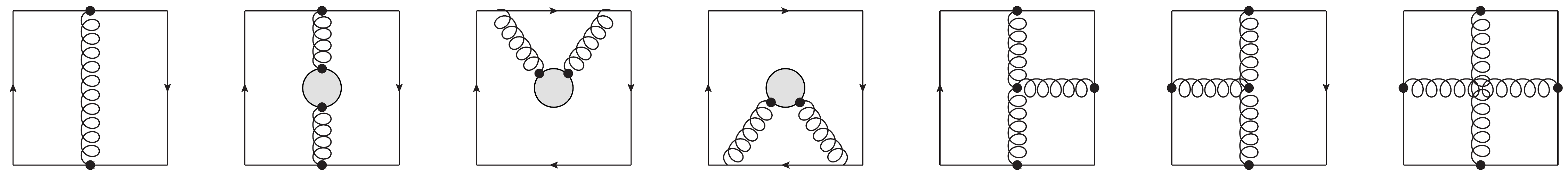}
 \caption{All relevant diagrams for $W_c$ at ${\cal O}(\alpha_s^2)$ in Coulomb gauge (time runs in the horizontal direction)}
 \label{calculation}
\end{figure}

The advantage of the exponentiation theorem is, that it is in general easier to calculate colour-connected coefficients instead of calculating reducible diagrams. So in the following we will need to consider only 2PI diagrams, and furthermore only those 2PI diagrams which do not cancel through cyclicity. For calculations, which have been regularized in dimensional regularization with $d=3-2\varepsilon$, Coulomb gauge turns out to be very advantageous, because in this gauge several of the remaining diagrams simply give zero. The relevant diagrams at ${\cal O}(\alpha_s^2)$ are given in figure \ref{calculation}. After removing the UV divergence from the gluon self energy through charge renormalization (in $\overline{\mathrm{MS}}$, with $1/\bar{\varepsilon}=1/\varepsilon-\gamma_E+\ln4\pi$) the result reads
\begin{align}
 \ln W_c=&\frac{C_F\alpha_s}{rT}\Biggl\{1+\frac{\alpha_s}{4\pi}\left[\left(\frac{31}{9}C_A-\frac{20}{9}T_Fn_f\right)+\left(\frac{11}{3}C_A-\frac{4}{3}T_Fn_f\right)\ln\mu r\right]\Biggr.\notag\\
&\hspace{41pt}+\Biggl.4\pi\alpha_s r \int\frac{\d^3k}{(2\pi)^3}\frac{\left(e^{i\r\cdot\k}-1\right)}{k^4}\left(-\Pi_{00}^{(T)}(0,\k)\right)\Biggr\}+C_FC_A\alpha_s^2\notag\\
&+\frac{C_FC_A\alpha_s^2}{\pi rT}\left(\frac{1}{\bar{\varepsilon}}+1+2\gamma_E-2\ln2+\ln\mu^2r^2+2\sum_{n=1}^\infty\frac{(-1)^n\zeta(2n)}{n(4n^2-1)}(rT)^{2n}\right)+{\cal O}(g^5)\,.
\label{result}
\end{align}
The square brackets in the first line come from the vacuum part of the gluon self energy (e.g.~\cite{Andrasi:2003zf}), the second line starts with the contribution from the thermal part~\cite{Heinz:1986kz,Kapusta:2006pm}, whose integral is not performed analytically but is UV and IR finite. The other term in the second line comes from the diagrams with a 3-gluon vertex. The last line gives the contribution from the last diagram in figure~\ref{calculation}, where the series expansion of the thermal part is valid for $rT\leq1$. This is the only divergent contribution at this order in $\alpha_s$. The integration for the gluon propagator connecting the two quark lines in this diagram has been carried out in 3 dimensions. If this integral had been computed in $d$ dimensions, then its ${\cal O}(\varepsilon)$ part would give a finite contribution when multiplied by the $1/\bar{\varepsilon}$ term in eq.~\eqref{result}. But after the divergence is subtracted as shown in the following section, this finite term will also be removed, so it is not necessary to display it here.

We have implicitly assumed the hierarchy of scales $1/r\gg T\gg m_D\gg\alpha_s/r$ here, and give only the contribution from the highest scale $1/r$. This discussion deals with UV divergences, so the low-energy behaviour of finite temperature QCD is not relevant. The general discussion of UV divergences in the following section, however, will not depend on this hierarchy of scales. The result~\eqref{result} agrees with the appropriate limit of the calculation found in~\cite{Burnier:2009bk}.

This result shows the peculiar difference between $W_c$ and a usual rectangular Wilson loop with cusps regarding UV divergences. There is no cusp divergence at ${\cal O}(\alpha_s)$ (it has cancelled because of cyclicity), instead the divergence at ${\cal O}(\alpha_s^2)$ is not a constant term, but multiplies a function of $r$ and $T$. It can therefore not be removed through a multiplicative constant to $W_c$, which would be an additive constant to eq.~\eqref{result} because of the logarithm. The reason for this will become apparent in the next section.

\section{UV divergences in loop functions}
\label{3}

Here we will first discuss the origin of UV divergences in loop functions, i.e.~expectation values of traced Wilson loops, and then apply the findings to the cyclic Wilson loop. This first part is based on the analysis found in \cite{Dotsenko:1979wb,Brandt:1981kf}. In position space, UV divergences arise when vertices are contracted at one point. In the case of internal vertices this leads to the usual UV divergences removed by charge renormalization. For loop functions it is also possible to contract line vertices, giving rise to a new sort of UV divergences. Depending on where and how many of these line vertices are contracted, we can further distinguish different types of UV divergences. The superficial degree of divergence for such a contraction is given by
\begin{equation}
 \omega_{\mathrm{smooth}}=1-N_{\mathrm{ex}}\,\hspace{50pt}\omega_{\mathrm{singular}}=-N_{\mathrm{ex}}\,.
\end{equation}
$N_{\mathrm{ex}}$ denotes the number of external lines, i.e. propagators leading to vertices which are not at the contraction point. The subscript refers to the point of contraction, where the contour can be smooth, have an angle or intersect itself. The latter two cases are called singular points. Then there are three possibilities:
\textit{(a)} line vertices are contracted at a smooth point without external lines, which gives a linear divergence, \textit{(b)} line vertices are contracted at a smooth point with one external line, which gives a logarithmic divergence, \textit{(c)} line vertices are contracted at a singular point without external lines, which also gives a logarithmic divergence.

\begin{figure}
 \includegraphics[width=\linewidth]{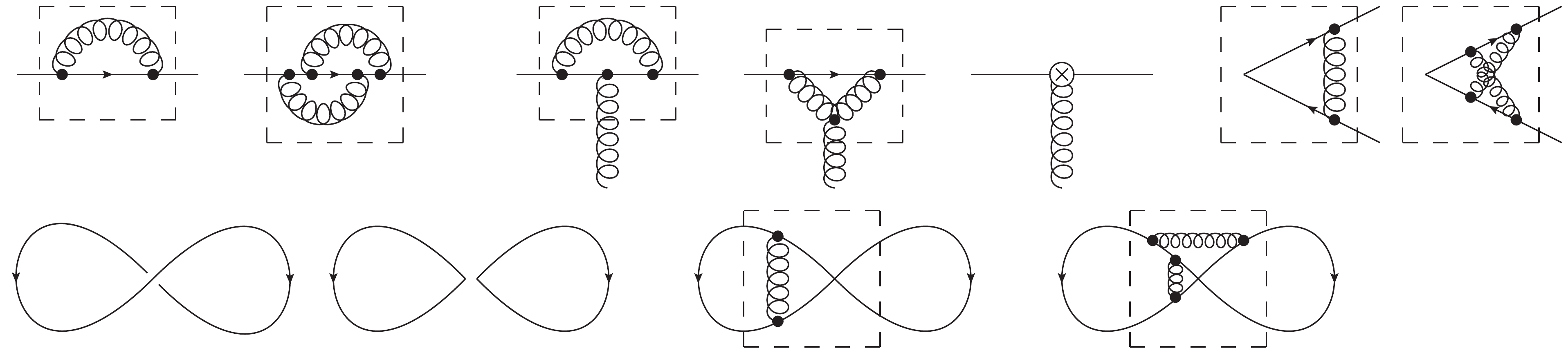}
 \caption{Illustrations for the different types of loop function divergences.}
 \label{types}
\end{figure}

Linear divergences, like those in \textit{(a)}, are automatically removed in dimensional regularization. They are therefore not relevant to this discussion. In other regularizations they may be absorbed in a ``loop mass'' term, which is proportional to the length of the contour.

The logarithmic divergences \textit{(b)} will be called \textit{line vertex divergences}, because they give in effect a correction to the line vertex of the external propagator and will be removed by a line vertex counterterm. This counterterm comes in through the factor $ig$ at each line vertex and is automatically included in the charge renormalization procedure. This was shown in \cite{Dotsenko:1979wb}. In the result \eqref{result} presented above charge renormalization has already been carried out, so line vertex divergences cannot be responsible for the divergence found.

The logarithmic divergences \textit{(c)} will be called \textit{cusp} or \textit{intersection divergences}, depending on the nature of the singular point. It was shown in \cite{Brandt:1981kf} that cusp divergences factorize out of loop functions and can therefore be removed through a multiplicative constant. This constant only depends on the cusp angle. Intersection divergences were also considered in \cite{Brandt:1981kf}, where the authors showed that loop functions with intersections mix under renormalization with other loop functions which have identical contours apart from a different path ordering prescription at the intersection. So for each intersection point there is a renormalization matrix mixing these associated loop functions. The renormalization matrices depend only on the angles at the respective intersection.

All four cases are illustrated in fig.~\ref{types}. The dashed boxes mean that all vertices inside the box are contracted at a point on the contour. If the box contains a singular point, then this is implied to be the contraction point. The first two diagrams are examples for linear divergences. The next two lead to a line vertex divergence which is removed through the counterterm in the following diagram. The last two diagrams in the first row are examples of cusp divergences. The dependence of the divergence on the cusp angle is introduced through diagrams like these, similar diagrams with all line vertices before or after the cusp also contribute to the cusp divergence but do not depend on the cusp angle. The second line shows the contours of two associated loop functions with a self-intersection, which have to be renormalized together. The contours are identical everywhere except for the intersection. If we denote the left and right loops as $W_L$ and $W_R$ respectively, then the first loop function is given by $\langle\Tr[W_LW_R]\rangle$ and the second by $\langle\Tr[W_L]\Tr[W_R]\rangle$. The last two diagrams are examples for intersection divergences. The first one of these appears in both loop functions and is identical to the contribution of a corresponding cusp divergence. The second example only contributes in the case of the first contour. It is diagrams like this which introduce the dependence on all intersection angles and necessitate the mixing of associated loop functions in order to renormalize them.

Turning back to the cyclic Wilson loop, due to the periodic boundary conditions the singular points are no longer cusps but intersections, because the two strings run along the same points in Euclidean space-time. In principle all points on the strings are intersections, but it can be shown that intersections where all angles are either $0$ or $\pi$ are not divergent, so only the two string endpoints need to be considered as intersections. The associated loop functions are given by $\left\langle\Tr\left[W(\r,0)P^\dagger(0,T)W(0,\r)\right]\Tr\left[P(\r,T)\right]\right\rangle$ and $\left\langle\Tr\left[P^\dagger(0,T)\right]\Tr\left[W(0,\r)P(\r,T)W(\r,0)\right]\right\rangle$ for choosing a different path ordering at the upper or lower intersection respectively, and for choosing a different path ordering at both intersections by $\left\langle\Tr\left[P^\dagger(0,T)\right]\Tr\left[W(\r,0)W(0,\r)\right]\Tr\left[P(\r,T)\right]\right\rangle$. In all three cases the string operators cancel, so every associated loop function corresponds to the Polyakov loop correlator.

So cyclic Wilson loop and Polyakov loop correlator are renormalized together by a $2\times2$ matrix. Observing that the Polyakov loop correlator is already finite after charge renormalization and that the angles at both intersection points are identical is enough to derive the following form of this renormalization matrix:
\begin{equation}
 \left(\begin{array}{c} W_c^{(R)} \\ P_c^{(R)} \end{array}\right)=\left(\begin{array}{cc} Z & 1-Z \\ 0 & 1 \end{array}\right)\left(\begin{array}{c} W_c \\ P_c \end{array}\right)\,.
\label{ren}
\end{equation}

By rearranging this equation we see that $W_c-P_c$ is multiplicatively renormalizable, while $W_c$ alone is not. Using eq.~\eqref{result} and the fact that $P_c=1+{\cal O}(g^3)$ (e.g.~\cite{McLerran:1981pb,Brambilla:2010xn}), we can determine the renormalization constant at first order to be $Z=1-\frac{C_A\alpha_s}{\pi\bar{\varepsilon}}+{\cal O}(\alpha_s^2)$. The renormalized Wilson loop is then given by eq.~\eqref{result} with the only modification that the $1/\bar{\varepsilon}$ term is absent.

\section{Test of the renormalization formula at ${\cal O}(\alpha_s^3)$}
\label{4}

In order to provide a test for this renormalization formula \eqref{ren}, we will discuss the intersection divergences arising at ${\cal O}(\alpha_s^3)$. Since the first order term of $Z=1+\sum_n Z_n\alpha_s^n$ has already been determined by the ${\cal O}(\alpha_s^2)$ calculation, $Z_1=-\frac{C_A}{\pi\bar{\varepsilon}}$, all contributions to the renormalized cyclic Wilson loop at ${\cal O}(\alpha_s^3)$ involving $Z_1$ must cancel in order for eq.~\eqref{ren} to be valid.

We will start by considering which diagrams can in principle contribute to the intersection divergence. The restriction to 2PI diagrams which do not cancel through cyclicity leaves only two possibilities for the remaining diagrams: either they have all their line vertices on the same quark line and none on the strings, or they have some line vertices on both quark lines. In the first case, these diagrams contribute equally to the Polyakov loop, and since the Polyakov loop is finite after charge renormalization, they cannot contribute to the intersection divergence of the cyclic Wilson loop. So divergent diagrams need to have at least one gluon connecting one quark line to the other. As explained in the previous section, intersection divergences arise when line vertices are contracted at the intersection without external lines. The gluon(s) connecting the quark lines cannot be contracted at any of the intersections, so there can be no connections between these gluons and the line vertices contracted at the intersection, otherwise there would be external lines.

So we can think of the intersection divergent diagrams as having two components. One consists of the gluons connecting the quark lines, which by themselves would be a finite diagram. We will call this the \textit{base diagram}. The other component consists of additional gluons added to the base, in such a way that they can be contracted at the intersection without external lines. These additional gluons need to have line vertices to the left and the right of the ones of the base diagram, otherwise this diagram would be reducible.

\begin{figure}
 \includegraphics[width=\linewidth]{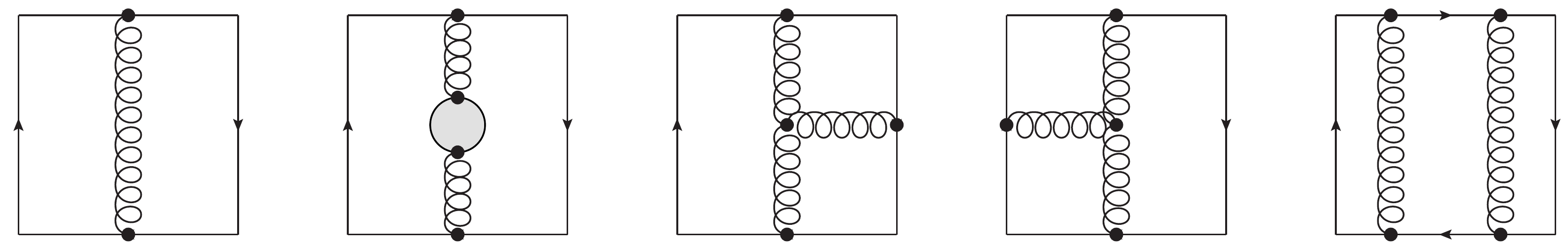}
 \caption{A list of all relevant bases up to ${\cal O}(\alpha_s^2)$ in Coulomb gauge. The last diagram is not 2PI, but after adding a gluon to make it intersection divergent it will be 2PI.}
 \label{bases}
\end{figure}

So in order to find all intersection divergent diagrams at ${\cal O}(\alpha_s^3)$ we just need to take all lower order bases and add gluons that can be contracted at the intersection. All relevant bases are shown in fig.~\ref{bases}. Again we will use Coulomb gauge for simplicity, otherwise there would be more diagrams to be considered. A treatment in more general gauges can be found in ref.~\cite{Berwein:2012mw}.

\begin{figure}
 \includegraphics[width=\linewidth]{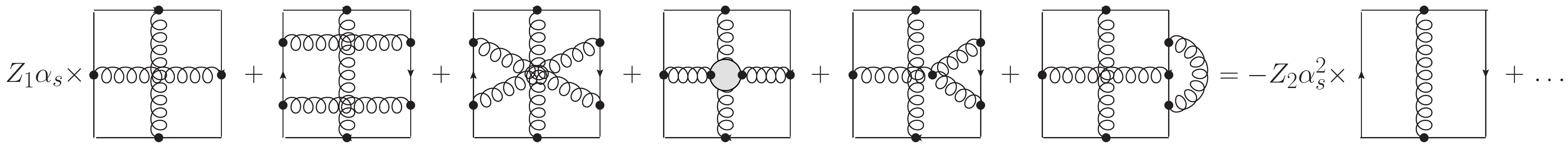}
 \caption{A sample of diagrams that determine the value of $Z_2$. The dots stand for remaining finite terms.}
 \label{Z2}
\end{figure}

Fig.~\ref{Z2} shows intersection divergent diagrams with the first diagram of fig.~\ref{bases} as base. Not all divergent diagrams with this base are displayed, but the remaining ones can be obtained from these by taking the mirror image or moving some line vertices from the strings to a quark line (but not beyond the line vertices of the base). The sum in fig.~\ref{Z2} is supposed to include all of them. The divergences are proportional to $\alpha_s^2$, so they have to be cancelled by $Z_2$. By calculating these diagrams, which we will not do here, one can determine the value of $Z_2$.

\begin{figure}
 \includegraphics[width=\linewidth]{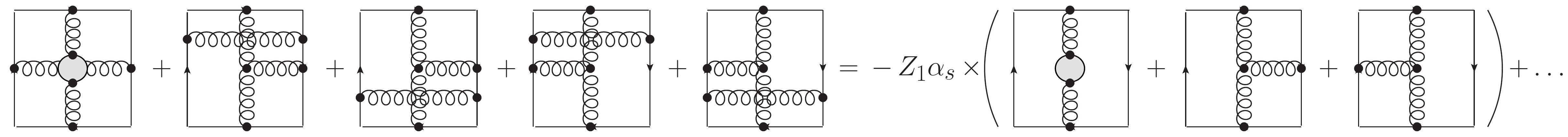}
 \caption{Intersection divergent diagrams which are renormalized similarly to the ${\cal O}(\alpha_s^2)$ divergence. Again, the dots stand for remaining finite terms.}
 \label{cancel}
\end{figure}

The other bases in fig.~\ref{bases} all need one additional gluon connecting the strings to become intersection divergent ${\cal O}(\alpha_s^3)$ diagrams. Except for the last one, the cancellation of their divergences is very similar to the case at ${\cal O}(\alpha_s^2)$. The added gluon gives the same divergent factor multiplying the base diagram and the divergence gets cancelled by $Z_1$ times the base. This is illustrated in fig.~\ref{cancel}.

\begin{figure}
 \includegraphics[width=\linewidth]{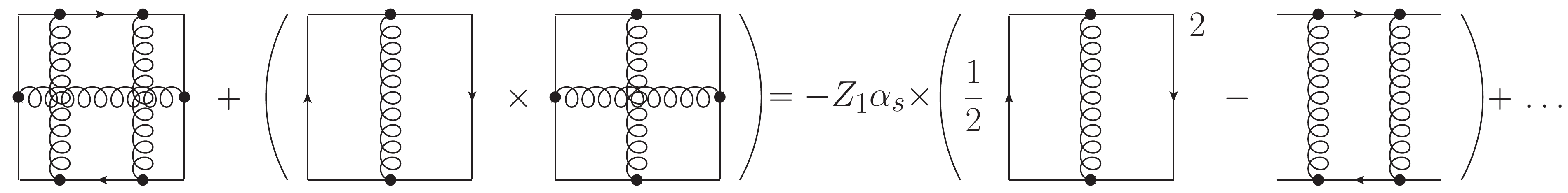}
 \caption{The cancellation of the remaining divergences in $W_c-P_c$ at ${\cal O}(\alpha_s^3)$ involves the Polyakov loop correlator.}
 \label{withP}
\end{figure}

The cancellation of divergences for diagrams with the last base in fig.~\ref{bases} is more complicated. An illustration is given in fig.~\ref{withP}. Since the sum over the diagrams shown in \ref{calculation} gives the logarithm of the cyclic Wilson loop, we need to re-expand the exponential when we calculate $W_c-P_c$. This is where the product of diagrams on the left-hand side and the square of the first diagram on the right-hand side come from. The last diagram in fig.~\ref{withP} comes from the Polyakov loop correlator, whose value at ${\cal O}(\alpha_s^2)$ is given by $P_c=1-(C_F^2-\frac{1}{2}C_FC_A)\frac{\alpha_s^2}{2r^2T^2}+2\delta P+{\cal O}(g^5)$ (e.g.~\cite{McLerran:1981pb,Brambilla:2010xn}). The part $\delta P$ comes from diagrams common to both $W_c$ and $P_c$, like the third and fourth diagram of fig.~\ref{calculation}, and therefore cancels in $W_c-P_c$. The relation shown in fig.~\ref{withP} in particular confirms the renormalization formula $W_c^{(R)}-P_c=Z(W_c-P_c)$, since here also the first non-trivial expansion term of $P_c$ plays an important role in the cancellation of divergences. This exhausts all possibilities for intersection divergent diagrams at ${\cal O}(\alpha_s^3)$ in Coulomb gauge.

\section{Conclusions}
\label{5}

The cyclic Wilson loop has been introduced and calculated in perturbation theory at ${\cal O}(\alpha_s^2)$. Using a general study of the divergences of loop functions we have analyzed the divergences of the cyclic Wilson loop and found that they do not originate from cusps but from intersections. Accordingly the cyclic Wilson loop cannot be renormalized by itself, one rather has to mix it with the Polyakov loop correlator to obtain a finite quantity. As a result of this approach we find that $W_c-P_c$ is in fact a multiplicatively renormalizable operator. This relation is confirmed by a study of the intersection divergent diagrams at ${\cal O}(\alpha_s^3)$. Extensions of this work can be found in \cite{Berwein:2012mw}.

\acknowledgments

I thank Nora Brambilla, Jacopo Ghiglieri and Antonio Vairo for their collaboration on the work presented here. Part of this work was done at Kyoto University and supported by the Global Cluster of Excellence (GCOE); I thank Hideo Suganuma and his group for their warm hospitality. This research was supported by the DFG cluster of excellence ``Origin and Structure of the Universe''.

\end{document}